# Tungsten Nanowire Based Hyperbolic Metamaterial Emitters for Near-field Thermophotovoltaic Applications

Jui-Yung Chang, Yue Yang, and Liping Wang[*]

School for Engineering of Matter, Transport & Energy
Arizona State University, Tempe, AZ 85287, USA

**ABSTRACT**

Recently, near-field radiative heat transfer enhancement across nanometer vacuum gaps has been intensively studied between two hyperbolic metamaterials (HMMs) due to unlimited wavevectors and high photonic density of state. In this work, we theoretically analyze the energy conversion performance of a thermophotovoltaic (TPV) cell made of $In_{0.2}Ga_{0.8}Sb$ when paired with a HMM emitter composed of tungsten nanowire arrays embedded in $Al_2O_3$ host at nanometer vacuum gaps. Fluctuational electrodynamics integrated with effective medium theory and anisotropic thin-film optics is used to calculate the near-field radiative heat transfer. It is found that the spectral radiative energy is enhanced by the epsilon-near-zero and hyperbolic modes at different polarizations. As a result, the power output from a semi-infinite TPV cell is improved by 1.85 times with the nanowire HMM emitter over that with a plain tungsten emitter at a vacuum gap of 10 nm. Moreover, by using a thin TPV cell with 10 μm thickness, the conversion efficiency can be greatly improved from 19.5% to 31.5% without affecting the power generation, due to the total internal reflection occurring at the bottom cell interface that minimizes the sub-bandgap spectral radiative energy. Furthermore, the effects of a TPV cell and a nanowire emitter with finite thicknesses are also studied. The result shows that the maximum efficiency of 32.2% is achieved with an optimal cell thickness of 3 μm while the nanowire HMM emitter should be thick enough to be opaque. The fundamental understanding and insights obtained here will facilitate the design and application of novel materials in enhancing near-field TPV energy conversion.

[*]Corresponding author. Email: liping.wang@asu.edu, Phone: 1-480-727-8615

## 1. INTRODUCTION

Near-field thermal radiation, which has been demonstrated to exceed the Planck's blackbody limit by orders of magnitude [1-5], has been proposed in the thermophotovoltaic (TPV) application due to the enhanced radiative heat transfer and thereby the power generation [6-11]. Coupled surface plasmon polariton (SPP) and surface phonon polariton (SPhP) have been proven to be the dominating resonance enhancement mechanisms in near-field thermal radiation [1, 2]. SPP or SPhP coupling across the vacuum gap requires similar materials for the emitter and receiver that support SPP or SPhP at close frequencies such that the coupling strength is strong and resulting near-field heat transfer could be greatly enhanced. However, the TPV receivers are usually photovoltaic cells made of semiconductors, while the TPV emitters require high-temperature materials such as tungsten. Therefore, it has been a daunting challenge to enhance near-field TPV performance with coupled SPPs across the nanometer vacuum gap due to the inherent material mismatch. Recently, graphene has been proposed to enhance the performance of TPV cells [12, 13]. Other physical mechanisms are highly desired for improving near-field TPV energy conversion.

Hyperbolic metamaterials (HMMs), such as metallic nanowires embedded in dielectric materials [14, 15], have shown the potential in numerous applications for far-field radiation [16-21], due to the unlimited length of propagating wave vector (i.e., high-$k$ mode) and broadband enhancement on photonic density of state (PDOS) [16, 17, 22-28] in addition to the epsilon-near-zero (ENZ) and epsilon-near-pole (ENP) modes [29]. Recently, nanowire-based HMMs have also been intensively studied for enhancing near-field thermal radiation by several times or even orders of magnitude over that with bulk materials [15, 22-24, 30-34]. The broadband resonance-free enhancement with HMMs in near-field radiative transport suggests their great potential in improving near-field TPV energy conversion. However, most of previous studies considered near-field radiative transfer between two identical HMMs across the vacuum gap, while the near-field TPV emitter and cell are usually made of dissimilar materials. A detailed analysis and understanding of a HMM emitter on the conversion performance of a near-field TPV cell is still lacking.

The present work studies the effect of a HMM emitter made of a tungsten nanowire array embedded in aluminum oxide ($Al_2O_3$) on the near-field TPV energy conversion. The influence of the volumetric filling ratio of the HMM emitter and the vacuum gap distance on near-field radiative transfer will be thoroughly investigated on the radiative transport, power generation and the conversion efficiency. Possible enhancement on the near-field radiation due to the ENP and hyperbolic modes of the HMM emitter will be explored. Moreover, the effect of a thin-film TPV cell with finite thickness will be studied aiming to further improve the conversion efficiency. Finally, the thickness effects of both a thin film TPV cell and a HMM emitter with vacuum substrate on both power output and conversion efficiency will be explored to optimize the cell performance.

## 2. THEORETICAL METHODS

### 2.1. Effective Medium Theory

Figure 1 shows the configuration of the near-field TPV system to be investigated in this study, which consists of a HMM emitter (Layer 1) made of tungsten nanowire arrays embedded in the $Al_2O_3$ host, and a TPV receiver (Layer 3) made of $In_{0.2}Ga_{0.8}Sb$ [35] separated by a vacuum gap (Layer 2) with distance *d*. The HMM emitter and the TPV receiver respectively have a finite thickness *h* and *t*, and vacuum is assumed to be the substrates (Layer 0 and Layer 4). The volumetric filling ratio of the nanowire array is $f = \pi D^2/(4a^2)$, where $a$ is the period and $D$ is the nanowire diameter, respectively.

Since the HMM emitter is an inhomogeneous medium made of tungsten and $Al_2O_3$, effective medium theory (EMT), which is a homogenization approach based on field average, is employed here to obtain effective dielectric functions of the emitter by approximating it as a homogeneous medium. Maxwell-Garnett (MG) method, which approximates the effective dielectric properties by treating one constituent as the embedded filler (i.e., nanowire) and all other constituents as host (i.e., $Al_2O_3$), is applied in this work. Note that the MG approximation assumes fillers independent with each other, which is an appropriate assumption for small filling ratios. Therefore, only the HMM emitter with $f < 0.5$ is investigated in this study. By setting depolarization factors $g_\perp = 0$ (when **E** field is perpendicular to the wire axis) and $g_\parallel = 0.5$ (when **E** field is parallel with the wire axis) for nanowires with high aspect ratios $h/D > 20$ [36], the MG theory can be expressed as [37, 38]:

$$\varepsilon_{\parallel,\text{eff}} = \varepsilon_d \frac{(\varepsilon_m + \varepsilon_d) + f(\varepsilon_m - \varepsilon_d)}{(\varepsilon_m + \varepsilon_d) - f(\varepsilon_m - \varepsilon_d)} \quad (1)$$

and 
$$\varepsilon_{\perp,\text{eff}} = \varepsilon_d + f(\varepsilon_m - \varepsilon_d) \quad (2)$$

where subscripts m and d represent tungsten (i.e., filler material) and $Al_2O_3$ (i.e., host material), respectively. The subscript $\parallel$ and $\perp$ denote the ordinary and extraordinary component of dielectric function. Clearly, the effective properties are highly dependent on the filling ratio, filler and host materials. Note that the EMT is applicable only if the feature size of the nanowire array is much smaller than thermal wavelength, e.g., $a/\lambda_{\text{th}} < 10$ [39]. The optical properties of tungsten and alumina are obtained from Palik's tabular data and assumed to be independent of temperature [40].

## 2.2. Hyperbolic Dielectric Behavior of Tungsten Nanowire Arrays

After the effective dielectric functions of the HMM emitter are obtained by the MG theory, the emitter can be treated as a homogeneous but uniaxial medium with a dielectric function in a tensor form as:

$$\bar{\bar{\varepsilon}} = \begin{bmatrix} \varepsilon_{\parallel,\text{eff}} & 0 & 0 \\ 0 & \varepsilon_{\parallel,\text{eff}} & 0 \\ 0 & 0 & \varepsilon_{\perp,\text{eff}} \end{bmatrix} \qquad (3)$$

Therefore, with certain conditions the nanowire emitter can exhibit hyperboloid shapes on the isofrequency relation for TM waves:

$$\frac{k_x^2 + k_y^2}{\varepsilon_{\perp,\text{eff}}} + \frac{k_z^2}{\varepsilon_{\parallel,\text{eff}}} = \left(\frac{\omega}{c}\right)^2 \qquad (4)$$

where $\omega$ is the angular frequency of incident wave, $c$ is the speed of light in vacuum, and $k_x$, $k_y$, and $k_z$ are the wavevector components on x, y, and z directions, respectively. Hyperboloid shapes only appear when $\varepsilon_{\perp,\text{eff}}$ and $\varepsilon_{\parallel,\text{eff}}$ have opposite signs, i.e., $\varepsilon_{\parallel,\text{eff}} > 0$ and $\varepsilon_{\perp,\text{eff}} < 0$ (denoted as type I HMM), or $\varepsilon_{\parallel,\text{eff}} < 0$ and $\varepsilon_{\perp,\text{eff}} > 0$ (denoted as type II HMM). Both types of HMM possess unlimited length of wavevector since hyperboloid shapes are unbounded, and thus can support high-$k$ mode, which allows propagating waves with large wavevectors [14]. Furthermore, this high-$k$ mode also supports a broadband singularity in photonic density of states (PDOS) [27, 28]. Both the large wavevector of propagating waves and the high PDOS inside HMMs are beneficial for enhancing near-field radiation heat transfer [14].

## 2.3. Near-Field Radiative Heat Transfer between a HMM Emitter and a TPV Cell Both with Finite Thicknesses

Fluctuational electrodynamics, which is based on the nature of random motion of charges or dipoles in media [41], is applied to calculate the near-field thermal radiation between the emitter and receiver. Recent work mainly focused on the near-field thermal radiation between isotropic thin films [42] or between semi-infinite uniaxial media [30]. Near-field radiation between two uniaxial thin films, however, is little studied. Here, we extend theoretical framework from Francoeur et al [43] on the near-field heat transfer between isotropic thin films to that between two uniaxial layers with finite thicknesses, by incorporating anisotropic wave optics. The net near-field heat flux between two uniaxial media with finite thicknesses at different temperatures ($T_1$ = 2000 K for emitter and $T_3$ = 300 K for receiver in this study) separated by a vacuum gap of $d$ can be expressed as [1, 32, 43]:

$$q'' = \int_0^\infty q''_\omega d\omega = \frac{1}{4\pi^2} \int_0^\infty d\omega \left[ \Theta(\omega, T_1) - \Theta(\omega, T_3) \right] \int_0^\infty \xi(\omega, \beta) \beta d\beta \tag{5}$$

where $\Theta(\omega, T) = \dfrac{\hbar\omega}{\exp(\hbar\omega/k_B T) - 1}$ represents the spectral mean energy of a Planck oscillator at temperature $T$. $\hbar$ and $k_B$ denote the reduced Planck constant and the Boltzmann constant, respectively. The energy transmission coefficient $\xi(\omega,\beta)$ has different expressions for propagating waves and evanescent waves as [43]:

$$\xi_{prop}(\omega, \beta) = \frac{\left(1 - |R_1^s|^2 - |T_1^s|^2\right)\left(1 - |R_3^s|^2 - |T_3^s|^2\right)}{\left|1 - R_1^s R_3^s e^{i2\gamma_2 d}\right|^2} + \frac{\left(1 - |R_1^p|^2 - |T_1^p|^2\right)\left(1 - |R_3^p|^2 - |T_3^p|^2\right)}{\left|1 - R_1^p R_3^p e^{i2\gamma_2 d}\right|^2} \tag{6a}$$

and

$$\xi_{evan}(\omega, \beta) = \frac{4\,\mathrm{Im}(R_1^s)\,\mathrm{Im}(R_3^s) e^{-2\mathrm{Im}(\gamma_2)d}}{\left|1 - R_1^s R_3^s e^{i2\gamma_2 d}\right|^2} + \frac{4\,\mathrm{Im}(R_1^p)\,\mathrm{Im}(R_3^p) e^{-2\mathrm{Im}(\gamma_2)d}}{\left|1 - R_1^p R_3^p e^{i2\gamma_2 d}\right|^2} \tag{6b}$$

where

$$R_j^\alpha = \frac{r_{j-1,j}^\alpha + r_{j,j+1}^\alpha e^{i2\gamma_j t_j}}{1 + r_{j-1,j}^\alpha r_{j,j+1}^\alpha e^{i2\gamma_j t_j}} \tag{7a}$$

and
$$T_j^\alpha = \frac{t_{j-1,j}^\alpha t_{j,j+1}^\alpha e^{i2\gamma_j t_j}}{1 + r_{j-1,j}^\alpha r_{j,j+1}^\alpha e^{i2\gamma_j t_j}} \quad (7b)$$

Here, the subscript number $j = 0, 1, 2, 3,$ or $4$ is the layer index shown in Fig. 1, which represents the vacuum substrate, emitter, vacuum gap, TPV receiver, and vacuum substrate, respectively. $\alpha = s$ or $p$ denotes the polarization states. $r_{j-1,j}^s = (\gamma_{j-1} - \gamma_j)/(\gamma_{j-1} + \gamma_j)$ and $r_{j-1,j}^p = (\varepsilon_j \gamma_{j-1} - \gamma_j)/(\varepsilon_j \gamma_{j-1} + \gamma_j)$ are the Fresnel reflection coefficients for $s$ and $p$ polarizations at the interface of layers $j-1$ and $j$, respectively. $\gamma_j$ is the z-component wavevector in an uniaxial layer $j$ and can be expressed differently for s and p polarizations [30, 34]:

$$\gamma_j^s = \sqrt{\varepsilon_{j\|}\omega^2/c^2 - \beta^2} \quad (8)$$

and
$$\gamma_j^p = \sqrt{\varepsilon_{j\|}\omega^2/c^2 - \frac{\varepsilon_{j\|}}{\varepsilon_{j\perp}}\beta^2} \quad (9)$$

where $\beta$ is the parallel component of wavevector $k_j$, and is identical in different layers due to the continuity boundary condition at the interfaces. When the layer is isotropic, $\gamma_j^p = \gamma_j^s$ is held with $\varepsilon_{j\|} = \varepsilon_{j\perp}$.

Note that when the thicknesses of emitter and receiver become infinitely large, the structure reduces to a three-layer configuration with a half-space emitter (Layer 1), vacuum (Layer 2), and a half-space receiver (Layer 3). In this case, $\xi(\omega,\beta)$ in Eq. (6) can be simplified through replacing the reflection coefficient of a thin film ($R_j^\alpha$) by that at an interface ($r_{j,j-1}^\alpha$ or $r_{j,j+1}^\alpha$) and setting $T_j^\alpha = 0$ [30]. In addition, the near-field radiative transfer in a four-layer thermal rectifier made of vacuum, a thin-film emitter, vacuum gap, and a half-space receiver has been investigated with a similar approach [44].

**2.4. TPV Power Generation and Conversion Efficiency**

When the photons with energy above the bandgap of the TPV cell, which is $E_g$ = 2.2 μm for $In_{0.2}Ga_{0.8}Sb$, are absorbed by the TPV receiver, electron-hole pairs could be generated and electricity could be produced with external loads. However, the photons with lower energy cannot excite electron hole pairs and result in low conversion efficiency. The near-field radiative transfer and the charge transport inside the TPV cell is a coupled problem, which has been theoretically modeled for multilayer isotropic media by calculating the charge density distribution due to the number of photons absorbed at different cell location [9]. In fact, the quantum efficiency in response to the near-field thermal radiation is different from that to a far-field source because of the inherent difference between near-field and far-field thermal radiation spectra. However, due to the limitation of the present theoretical model, which cannot predict the spectral energy absorbed at different depths by treating the TPV cell as a multilayer, the quantum efficiency to the near-field thermal radiation between the uniaxial HMM emitter and the TPV cell cannot be thus obtained accurately. Instead of assuming a perfect TPV cell with 100% quantum efficiency, which will overpredict the performance, we use the far-field internal quantum efficiency of $In_{0.2}Ga_{0.8}Sb$ [35] for evaluating the proposed near-field TPV system with a nanowire-based HMM emitter, aiming for more practical insights. For the sake of completeness, the methods for calculating the power generation and conversion efficiency are summarized here.

With the near-field spectral heat flux $q_\lambda(\lambda)$ calculated from Eq. (5) and the far-field internal quantum efficiency data $\eta_q(\lambda)$, the short-circuit current can be obtained as [9]:

$$J_{ph} = \frac{e}{hc} \int_0^\infty \eta_q(\lambda) q_\lambda(\lambda) \lambda d\lambda \tag{10}$$

The dark current can be calculated by [9]:

$$J_0 = e\left(\frac{n_{in}^2 D_h}{N_D \sqrt{\tau_h}} + \frac{n_{in}^2 D_e}{N_A \sqrt{\tau_e}}\right) \quad (11)$$

Here, $n_{in}$ is the intrinsic carrier concentration, $N_D$, and $N_A$ are the carrier concentration of donor and acceptor, respectively. $D_h$, $D_e$, $\tau_h$, and $\tau_e$ are the diffusion coefficient and relaxation time of holes and electrons, respectively. Moreover, the open-circuit voltage is given by:

$$V_{oc} = (k_B T / e) \ln(J_{ph} / J_0 + 1) \quad (12)$$

Therefore, the electrical power output can be expressed as [9]:

$$P_{El} = J_{ph} V_{oc} (1 - 1/y)[1 - \ln(y)/y] \quad (13)$$

where
$$y = \ln(J_{ph} / J_0) \quad (14)$$

Finally, by dividing the electrical power output with the total radiative power input, the conversion efficiency of the proposed near-field TPV system with a HMM emitter is [9]:

$$\eta = P_{El} / P_R \quad (15)$$

## 3. RESULTS AND DISCUSSION

### 3.1. Effective Dielectric Functions of the Hyperbolic Metamaterial Emitter

Figures 2(a) and 2(b) show the real part of effective dielectric functions for parallel and vertical components of the HMM emitter, respectively. Clearly the tungsten nanowire array behaves as the type I HMM within frequencies at which $\varepsilon_{\perp,eff}$ is negative while and $\varepsilon_{\parallel,eff}$ is always positive within the considered spectral regime. Furthermore, the filling ratio effect on effective dielectric functions can also be observed. Figure 2(a) illustrates that, with increasing filling ratios the ENP peak for parallel effective dielectric function becomes larger, which indicates stronger ENP enhancement under large filling ratios. On the other hand, Fig. 2(b) demonstrates that, the characteristic frequency at which the vertical effective dielectric function becomes negative will shift to higher frequencies with larger filling ratios. Therefore, since the parallel effective dielectric function is always positive under all filling ratios throughout the whole spectrum, higher filling ratio will result in type I HMM region shifting to higher frequency. Note that at the ENP region shown in Fig. 2(a), plain tungsten has metallic behavior which does not support ENP enhancement.

Figures 2(c) and 2(d) show the imaginary part of parallel and vertical effective dielectric functions of the HMM emitter, respectively. The effect of filling ratio on the ENP enhancement can also be observed in Fig. 2(c) with higher ENP peak. Similar effect of filling ratio can also be seen in Fig. 2(d) on the vertical effective dielectric function which becomes more lossy at higher filling ratio.

Note that all four figures characterize the effective dielectric functions of both parallel and vertical components of the tungsten nanowire based HMM for both s and p polarizations. For s polarization in which only the parallel component dielectric function takes effect, the underlying mechanism is the ENP mode, which would result in the reduction of impedance mismatch and thus higher energy transfer [29]. On the other hand, the HMM behavior will appear at p polarization. The angular frequency region and wavelength band of the type I HMM behavior can be classified by optical phase diagram with respect to different filling ratios as shown in Fig. 2(e). Clearly, the region of type I HMM extends to smaller wavelengths (i.e., higher frequencies) with larger filling ratios. Besides, the tungsten nanowire arrays behave as an effective dielectric with both positive $\varepsilon_{\parallel,\text{eff}}$ and $\varepsilon_{\perp,\text{eff}}$ outside of the HMM regime.

**3.2. TPV Performance with a Semi-infinite Nanowire Emitter and a Semi-infinite PV Cell**

Let us first consider a near-field TPV system made of a semi-infinite tungsten nanowire HMM emitter and a semi-infinite TPV cell. At a vacuum gap of $d = 10$ nm, the effect of filling ratio on the transmission coefficients, spectral heat flux, power output, and conversion efficiency will be studied subsequently, followed by the effect of vacuum gap distances on the TPV performance. All the calculations in this and following sections are completed with sufficient amount of data points to ensure the accuracy in numerical integration: 1000 data points in the frequency domain of $3.7 \times 10^{14}\,\text{rad/s} \leq \omega \leq 6.2 \times 10^{15}\,\text{rad/s}$, 1000 points for propagating waves with $0 \leq \beta c/\omega < 1$, and 10000 points for evanescent waves with $\beta c/\omega > 1$. The numerical error is checked to be less than 0.1% when the amounts of data points for angular frequency $\omega$ and dimensionless wavevector $\beta c/\omega$ are doubled.

*3.2.1. Transmission Coefficients at Different Polarizations.* Figures 3(a), 3(b), and 3(c) show the transmission coefficient $\xi(\omega,\beta)$ at s polarization with filling ratio $f = 0.1$, 0.3, and 0.5, respectively, when the vacuum gap is fixed at $d = 10$ nm. By comparison, the transmission coefficients in the shaded region are enhanced with more channels of heat transfer (i.e., $\beta c/\omega$) at larger filling ratios, which can be explained by the stronger ENP mode at the same frequency region shown in Figs. 2(a) and 2(c). Note that, only the parallel component of dielectric function of the HMM emitter $\varepsilon_{\parallel,\text{eff}}$ is involved with s-polarized waves. When the filling ratio increases, higher absorption loss (or emission) takes place at the ENP mode for $\varepsilon_{\parallel,\text{eff}}$, and results in larger radiative transfer or transmission coefficients from the emitter to the receiver.

Figures 3(d), 3(e), and 3(f) present the transmission coefficient $\xi(\omega,\beta)$ at p polarization with $f = 0.1$, 0.3, and 0.5, respectively. As the filling ratio increases, the shaded region with large

transmission coefficient at low frequencies gets broadened towards higher frequency, which matches well with the type I HMM region as shown in Fig. 2(e). Therefore, the enhancement in $\xi(\omega,\beta)$ at low-frequency region is due to the HMM behavior of the tungsten nanowire emitter. The enhancement at higher frequencies is due to the effective dielectric behavior of the nanowire emitter, in which the absorption loss becomes greater with larger imaginary parts of both $\varepsilon_{\parallel,\text{eff}}$ and $\varepsilon_{\perp,\text{eff}}$ with increasing filling ratio, as indicated by Figs. 2(c) and 2(d).

*3.2.2. Spectral Radiative Heat Flux.* The enhancement of transmission coefficients will result in the enhanced spectral heat flux, according to Eq. (5). Figures 4(a) and 4(b) show the spectral heat fluxes between the semi-infinite nanowire HMM emitter and semi-infinite TPV cell for s and p polarizations, respectively. At s polarization, a steady enhancement on the spectral heat flux appears with the increment of filling ratio. Furthermore, due to the narrow-band ENP enhancement on transmission coefficient $\xi(\omega,\beta)$ demonstrated by Figs. 3(a), 3(b), and 3(c), an enhancement within the ENP region also takes place with larger filling ratios. Note that, even though the spectral heat flux for s polarization rises with $f$ from 0.1 to 0.5, it does not necessarily indicate that the spectral heat flux for plain tungsten emitter (i.e., $f = 1$) will become larger than that with nanowire emitters, which in fact is lower than that of nanowires with filling ratio $f > 0.3$. This can be explained by Figs. 2(a) and 2(c) where the effective material property of nanowire emitters is not similar to plain tungsten for high filling ratios up to $f = 0.5$. Plain tungsten has negative real part of dielectric function at angular frequency $\omega < 2\times10^{15}$ rad/s, which indicates metallic behavior. On the other hand, tungsten nanowires possess positive real part of $\varepsilon_{\parallel,\text{eff}}$ in the same frequency region, and behave as lossy dielectrics. As a result, the spectral heat flux from the nanowire emitters is higher than that from plain tungsten at longer wavelengths, as shown in Fig. 4(a). Note that, the spectral peak with plain tungsten at 1.4 μm is due to its intrinsic bandgap absorption.

As shown in Fig. 4(b) for p polarization, the spectral heat flux distribution with the HMM emitters exceeds that with plain tungsten at wavelengths $\lambda > 1.6$ μm or so for all filling ratios. The enhanced spectral heat flux region matches well with the type I HMM region as shown in Fig. 2(e). The spectral peak also shifts slightly towards lower wavelengths with larger filling ratios, which agrees with the trend observed from the enhanced transmission coefficient region in Fig. 3 for p polarization at different $f$ values. Therefore, the enhanced spectral flux with HMM emitters at longer wavelengths is due to the hyperbolic behavior in tungsten nanowires, which do not exist in plain tungsten. Among different filling ratios, the difference in spectral heat flux is not as apparent as that for s polarization. Furthermore, the spectral heat flux drops abruptly at the wavelength $\lambda \approx 2$ μm for all filling ratios, which does not appear for s polarization. The abrupt reduction in $q(\lambda)$ at p polarization can be explained by the sudden drop of real part of $\varepsilon_{\perp,\text{eff}}$ in Fig. 2(b), or equivalently by the sharp increase of the imaginary part of $\varepsilon_{\perp,\text{eff}}$ in Fig. 2(d).

Figure 4(c) shows the overall spectral heat flux from both s and p polarizations for tungsten nanowire HMM emitters with different filling ratios. The maximum spectral heat flux could be as high as 4.5 MW/m²-μm around $\lambda = 1.5$ μm with $f = 0.5$. Note that the TPV cell made of $In_{0.2}Ga_{0.8}Sb$ has a bandgap of $E_g = 2.2$ μm. The spectral heat flux from the HMM emitters with $f > 0.3$ is larger than that with plain tungsten at the wavelengths above the cell bandgap $E_g$, suggesting the potential to enhance power generation and conversion efficiency by replacing plain tungsten emitter with tungsten nanowires. However, the spectral heat flux below the cell bandgap $E_g$ is also enhanced with nanowire HMM emitters, indicating more wasted long-wavelength photons which might decrease the conversion efficiency. In order to gain a better understanding on how the nanowire HMM emitters would impact the TPV performance, the power generation and overall photon-to-electricity conversion efficiency need to be evaluated.

*3.2.3. Power Output and Conversion Efficiency.* Figure 5(a) shows the normalized radiative power input and electrical power output of the TPV cell with HMM emitters at different filling ratios over that with plain tungsten at the vacuum gap $d = 10$ nm. Note that, the radiative heat flux and power generation with a plain tungsten emitter are 3.63 MW/m$^2$ and 0.87 MW/m$^2$ respectively, leading to a conversion efficiency of 23.9%. By replacing the plain tungsten emitter with tungsten nanowire HMMs, the electrical power output increases with larger filling ratios, and exceeds that with plain tungsten under all filling ratios. The enhancement in power output could reach about 1.85 times higher with $f = 0.5$. On the other hand, the total radiative heat flux from the emitters to the TPV cell is also amplified with nanowire HMMs, and increases with filling ratio. This is because the HMM emitter enhances the spectral energy not only above but also below the bandgap at longer wavelengths as the broadband enhancement shown in Fig. 4(c). As a result, the conversion efficiency with nanowire HMM emitters is around 19% ~ 19.5% with different filling ratio, which is about 4% less compared with that (23.9%) from a plain tungsten emitter as shown in Fig. 5(b), though the electrical power output is enhanced almost by two times.

*3.2.4. Effect of Vacuum Gap Distance.* The electrical power output and the TPV conversion efficiency are calculated at different vacuum gap distances for the nanowire HMM emitter with $f = 0.5$, as shown in Figs. 6(a) and 6(b), respectively. The power generation decreases monotonically when the HMM emitter and the TPV cell are further apart, due to the reduction of near-field radiative heat transfer. However, the conversion efficiency with the HMM emitter drops first from 19% to a minimum of 8.5% when the vacuum gap becomes larger from $d$ = 10 nm to 300 nm or so, and then rises up to 15% at a 1-μm vacuum gap. The plain tungsten exhibits similar behaviors in both power output and conversion efficiency but with smaller $P_{El}$ and a larger efficiency $\eta$ by 4% or so. The different dependence on the vacuum gap between the electrical power and conversion efficiency can be understood by the effect of vacuum gap on the spectral radiative heat flux, as only that above the cell bandgap would enhance the conversion efficiency. However, in order to make the nanowire HMM emitter more useful in enhancing the near-field TPV conversion efficiency over the plain tungsten, other approaches such as considering the TPV emitter and cell as thin films need to be further explored.

## 3.3. Performance Enhancement from Nanowire HMM Emitters and TPV Cells with Finite Thicknesses

In this section, both the HMM emitter and the TPV cell will be considered as thin films, and the effect of film thicknesses on the near-field radiative heat transfer and the TPV conversion efficiency will be studied, aiming to enhance the TPV performance with the tungsten nanowire HMM emitters. First, the TPV cell will be modeled as a thin film while the emitter is still semi-infinite in a 4-layer near-field TPV design. Then, the thickness effect of both thin TPV cell and a nanowire HMM emitter will be investigated via a 5-layer structure, while the substrate for both the TPV cell and the HMM emitter is taken to be vacuum for simplicity.

*3.3.1. Thin Film Effect of a TPV Cell with Finite Thickness.* The key to enhance the TPV conversion efficiency is to reduce the spectral radiative energy below the bandgap that is not useful to generate electron-hole pairs in the cell [11]. Here, the semi-infinite TPV cell is replaced by a free-standing thin layer with thickness $t = 10$ μm, while the nanowire HMM emitter is kept semi-infinite, forming a 4-layer near-field TPV system. The filling ratio of the HMM emitter is chosen to be $f = 0.5$ for comparison with the previous results from the semi-infinite TPV cell.

Figure 7(a) shows the spectral radiative heat flux from the 4-layer TPV system with a cell thickness of $t = 10$ μm. It can clearly be observed that the spectral heat flux at wavelengths smaller than the cell bandgap is almost the same with that from a semi-infinite cell shown in Fig. 4(c) with relative difference less than 1%, indicating that the spectral radiative energy above the cell bandgap is totally absorbed within the 10-μm TPV cell. However, the spectral radiative energy below the bandgap is surprisingly reduced to almost zero at longer wavelengths, which is actually due to the total internal reflection occurring at the bottom interface of the PV cell and vacuum substrate because the electromagnetic waves are incident from a dense medium (i.e., TPV cell) with a larger refractive index into vacuum. In fact, only the waves with spectral energy below the bandgap could reach the bottom cell interface and totally reflected back. The significantly reduced long-wavelength radiative energy absorbed by the TPV cell due to the internal total reflection could greatly enhance the TPV conversion efficiency.

The normalized radiative power input and electrical power output from the 4-layer TPV system with the nanowire HMM emitter to that with plain tungsten emitter, as shown in Fig. 7(b), clearly shows the desired drop of the radiative power by comparison with Fig. 5(a). Since the spectral radiative energy above the cell bandgap is almost identical, the electrical power generation is almost kept unaffected when the TPV cell becomes a thin layer of 10 μm in thickness. As a result, the TPV conversion efficiency, which is the ratio of power generation to the radiative power input, is improved as expected due to the minimization of long-wavelength radiative transfer below the bandgap. As shown in Fig. 7(c), the improved conversion efficiency of TPV system with a semi-infinite HMM emitter and a thin TPV cell could reach 31.5% or so regardless filling ratios, which is nearly a 4% increase to that (27.5%) from a thin TPV cell of 10 μm thick or even a 8% increase over that (23.9%) from a semi-infinite TPV cell both with a plain tungsten emitter.

*3.3.2. Thickness Effects of Both Thin TPV Cell and Thin HMM Emitter.* As the improved near-field TPV performance was demonstrated with a thin TPV cell of 10 μm in thickness, we now aim to gain a quantitative understanding in the cell thickness effect on the TPV conversion efficiency and the power generation, which is shown in Fig. 8(a). The parameters such as vacuum gap $d = 10$ nm and emitter filling ratio $f = 0.5$ are kept the same. As the TPV cell thickness decreases from $t = 10$ μm, the power output drops monotonically. However, the conversion efficiency slightly rises up to a maximum of 32.2% around $t = 3$ μm. Within this thickness range, the TPV cell is still thick enough to absorb most of the spectral energy above the cell bandgap but absorbs less spectral energy below the bandgap due to smaller cell thickness (or less material). As a result, the conversion efficiency is slightly increased. When the TPV cell thickness $t$ is further reduced to 10 nm, the conversion efficiency drops dramatically down to ~26%, which is apparently because the cell is not thick enough to absorb all the spectral energy above the bandgap. Therefore, the thickness effect indicates that, the TPV cell should be chosen to be thicker than 2 μm to achieve the optimal conversion efficiency.

It should be noted that the present conversion analysis on the cell thickness effect does not consider the actual influence to the charge transport and recombination, which would affect the quantum efficiency of the thin TPV cell at different thickness. In order to rigorously model the charge transport coupled with the near-field radiative transport, the spectral energy absorbed at different cell depth has to be calculated [9, 10]. However, because of the uniaxial nature of the nanowire emitter, the rigorous calculation requires a multilayer near-field radiative transfer model incorporated with anisotropic wave optics, which is currently not available. Therefore, the present study considers the TPV cell (i.e., $In_{0.2}Ga_{0.8}Sb$) with quantum efficiency from the far-field response and neglects the thickness effect on the charge transport, aiming to focus on the effect of near-field radiative transfer with HMM emitter on the near-field TPV efficiency. The rigorous modeling of the HMM-enhanced near-field TPV systems considering the coupled radiative and charge transports will be pursued as the future work.

Finally, the thickness effect of the tungsten nanowire emitter on the near-field TPV performance is considered by treating the nanowire array with a finite height $h$ on vacuum substrate, forming a 5-layer near-field TPV system with a thin TPV cell with thickness $t = 10$ μm. As shown in Fig. 8(b), the radiative and electrical power as well as the resulting conversion efficiency do not change much when the nanowire emitter thickness $h$ reduces from 10 μm to 1 μm, suggesting that 1-μm-long nanowires are thick enough to be opaque. When further reducing the nanowire height down to 200 nm or so, oscillation occurs in both the radiative and electrical power plots as a function of emitter thickness $h$, due to the wave interference inside the thin-film emitter as the thermal wavelength is comparable with the nanowire thickness. As a result, the conversion efficiency also oscillates with a 1% fluctuation but does not change much. However, when the tungsten nanowire height is further reduced to be $h < 200$ nm, the radiative power drops dramatically with a much faster rate than the electricity power generation. This can be explained by the less emitting materials as the ENP and the HMM behavior are responsible for the near-field radiative heat transfer enhancement, both of which are simply associated with the effective material properties of the tungsten nanowire arrays. Therefore, the TPV conversion efficiency decreases significantly from 31% at $h$ = 200 nm to 24% with a thin nanowire array of 10 nm in thickness.

## 4. CONCLUSION

In summary, the performance of a near-field TPV system with HMM emitter made of tungsten nanowire embedded in $Al_2O_3$ host has been thoroughly investigated. The filling ratio effect on the material properties of near-field radiative heat transfer was analyzed. The results show that a semi-infinite HMM emitter with $f$ = 0.5 can achieve 1.85 times (1.56 MW/m$^2$) of enhancement in electrical power output from semi-infinite TPV cell made of $In_{0.2}Ga_{0.8}Sb$ comparing to that with plain tungsten emitter. The mechanisms for radiative heat transfer enhancement have been illustrated as ENP for s polarization and hyperbolic modes for p polarization. However, it turned out that the conversion efficiency becomes lower because the spectral radiative energy below the cell bandgap is also enhanced with the HMM emitter. In order to improve the conversion efficiency, a thin-film TPV cell with $t$ = 10 μm is considered. Total internal reflection occurs at the bottom interface of TPV cell and thus minimizes the energy absorption below band gap. As a result, the conversion efficiency has been enhanced from 19.5% to 31.5% with the same amount of power output, compared with 27.5% with a plain tungsten emitter. Furthermore, with a TPV cell thickness of $t$ = 3 μm, the conversion efficiency can be further improved to 32.2% where less energy below the bandgap is absorbed, while the nanowire emitter is desired to be opaque. The results gained here would facilitate the practical design and demonstration of near-field TPV devices enhanced by novel metamaterial emitters.

**ACKNOWLEDGMENT**

The authors are grateful to the support from the New Faculty Startup Program at ASU.

**NOMENCLATURE**

| | |
|---|---|
| $a$ | period of structure, m |
| $c$ | light velocity in vacuum, $2.998 \times 10^8$ m s$^{-1}$ |
| $D$ | diameter of nanowire, m |
| $d$ | gap distance, m |
| $E$ | electrical field |
| $e$ | electric charge, C |
| $f$ | filling ratio |
| $g$ | depolarization factor |
| $h$ | Planck constant |
| $J_0$ | dark current, A    $J_{ph}$    short circuit current, A |
| $k_B$ | Boltzmann constant, J/K |
| $k_{x,y,z}$ | x, y, z component of wavevector |
| $P_{El}$ | electrical power, W |
| $P_R$ | radiative power, W |
| $q''$ | heat flux, W/m$^2$ |
| $q_\omega''$ | spectral heat flux, W/(m$^2$ rad/s) |
| $q_\lambda$ | spectral heat flux, W/(m$^2$ m) |
| $r$ | reflection coefficient |
| $T$ | temperature, K |
| $V_{oc}$ | open circuit voltage, V |

*Greek Symbols*

| | |
|---|---|
| $\beta$ | parallel component of wavevector |
| $\gamma$ | vertical component of wavevector |
| $\varepsilon$ | dielectric function |
| $\omega$ | angular frequency, rad/s |
| $\Theta$ | Planck oscillator |
| $\xi$ | transmission coefficient |
| $\hbar$ | reduced Planck constant |
| $\lambda$ | wavelength, m |
| $\eta$ | conversion efficiency |
| $\eta_q$ | quantum efficiency |

*Subscripts*

| | |
|---|---|
| 0, 1,2,3,4 | layer 0, 1, 2, 3, 4 |
| ⊥ | vertical component |
| ∥ | parallel component |
| A | acceptor |
| $Al_2O_3$ | aluminum oxide |
| D | donor |
| El | electrical |
| e | elctron |
| eff | effective medium |
| evan | evanescent wave |
| h | hole |
| i, j | medium i, j |
| in | intrinsic |
| prop | propagating wave |
| R | radiative |
| W | tungsten |
| λ | spectral |

*Superscripts*

| | |
|---|---|
| α | state of polarization |
| s | s polarization |
| p | p polarization |

**FIGURE CAPTIONS:**

Fig. 1  Schematic of a TPV cell with tungsten nanowire based HMM emitter and PV receiver of finite thickness. (5-layer structure: vacuum substrate, emitter, vacuum gap, receiver, vacuum substrate)

Fig. 2  Dielectric functions of tungsten nanowire HMM with respect to different filling ratio: (a) real part of $\varepsilon_\parallel$; (b) real part of $\varepsilon_\perp$; (c) imaginary part of $\varepsilon_\parallel$; (d) imaginary part of $\varepsilon_\perp$. (e) Optical phase diagram of tungsten nanowire array based emitter with respect to angular frequency and wavelength.

Fig. 3  $\xi$ function of TVP cell for s polarization with filling ratio of: (a) 0.1; (b) 0.3; (c) 0.5 and for p polarization with filling ratio of: (d) 0.1; (e) 0.3; (f) 0.5. (3-layer structure)

Fig. 4  Spectral heat flux of TPV cell with HMM emitter for: (a) s polarized wave; (b) p polarized wave; (c) overall wave. (3-layer structure)

Fig. 5  (a) The electrical power output and the radiative power input with HMM emitter normalized to the power input and output with plain tungsten emitter. (b) The conversion efficiency of TPV cell with HMM and plain tungsten emitter. (3-layer structure)

Fig. 6  The vacuum gap effect on: (a) power output; (b) conversion efficiency of TPV cell with HMM and plain tungsten emitter. (3-layer structure)

Fig. 7  (a) Overall spectral heat flux of TPV cell with HMM emitter. (b) The electrical power output and the radiative power input with HMM emitter normalized to the power input and output with plain tungsten emitter. (c) The conversion efficiency of TPV cell with HMM and plain tungsten emitter. (4-layer structure)

Fig. 8  The thickness effect of: (a) receiver; (b) emitter on power input, power output, and conversion efficiency of TPV cells with vacuum substrates and HMM emitter ($f = 0.5$). (5-layer structure)

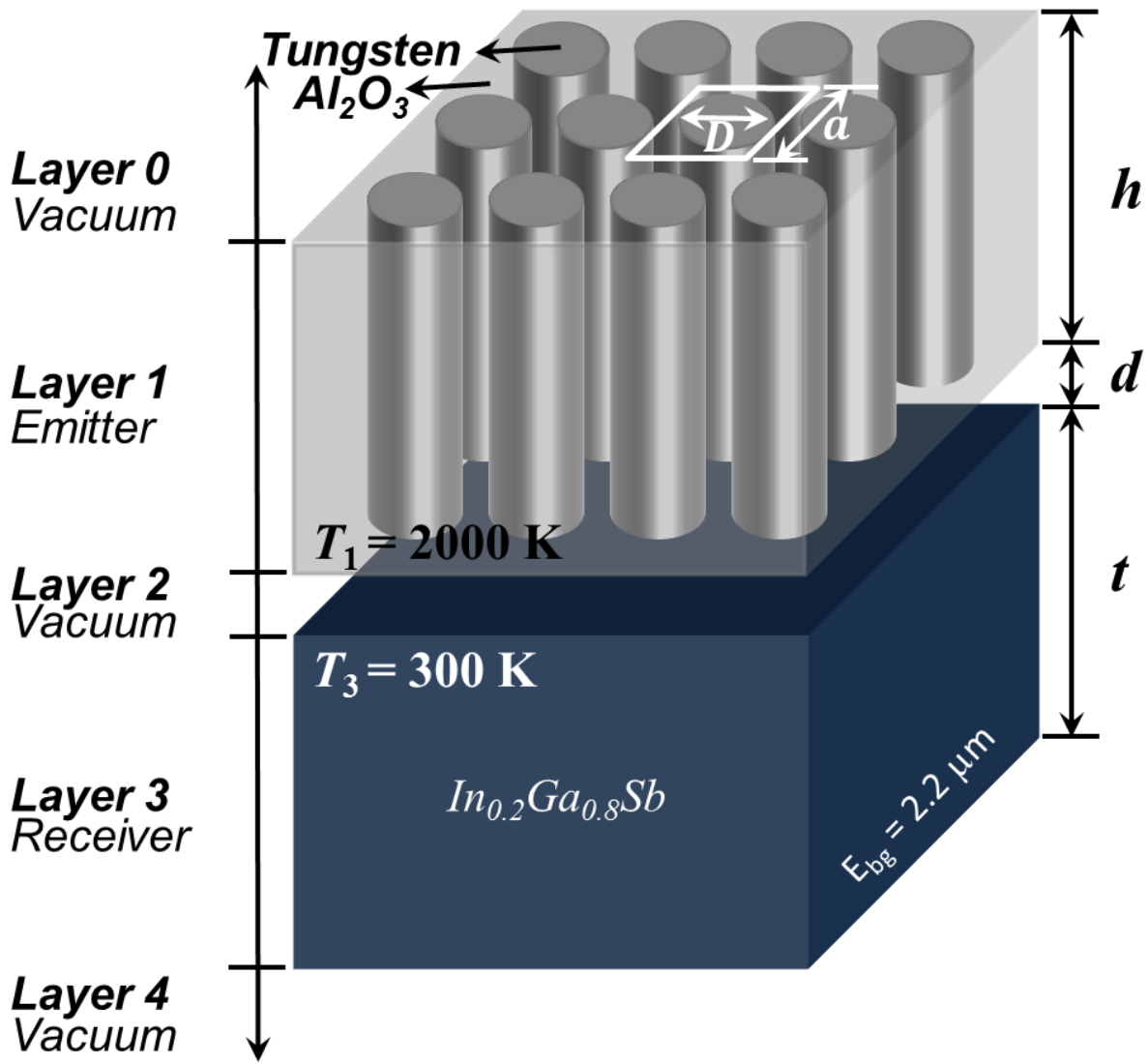

**Chang et al, Fig. 1**

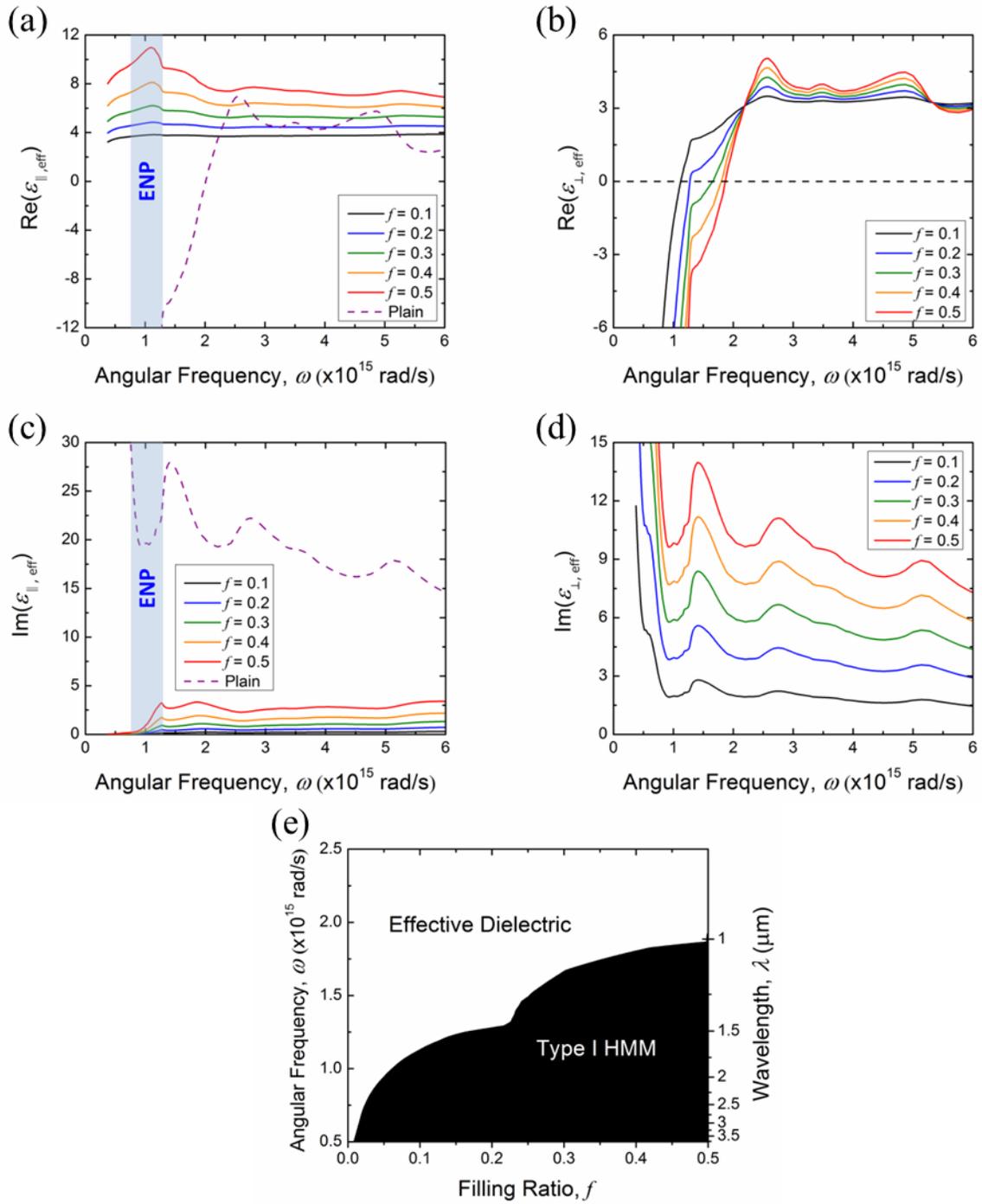

**Chang et al, Fig. 2**

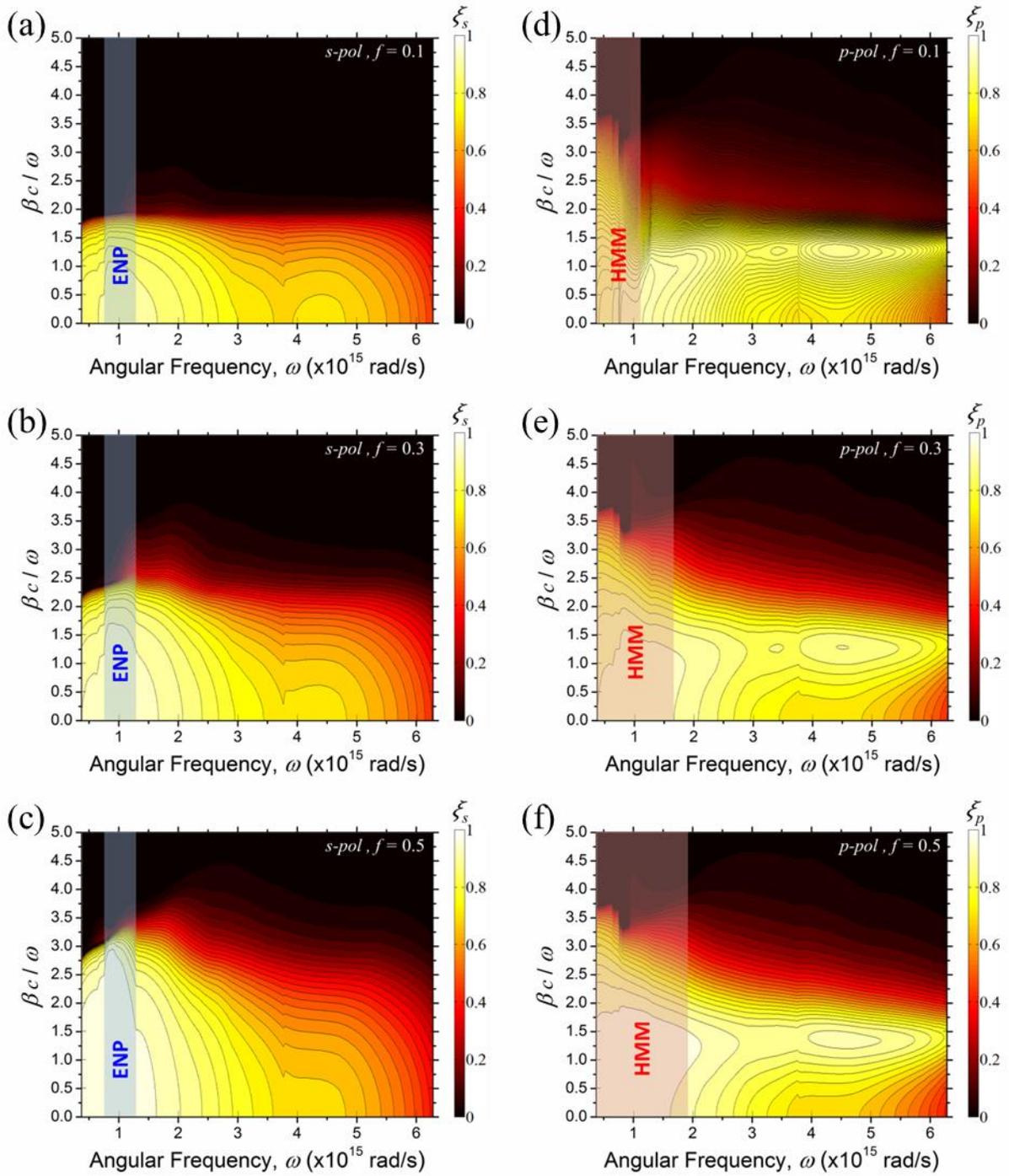

**Chang et al, Fig. 3**

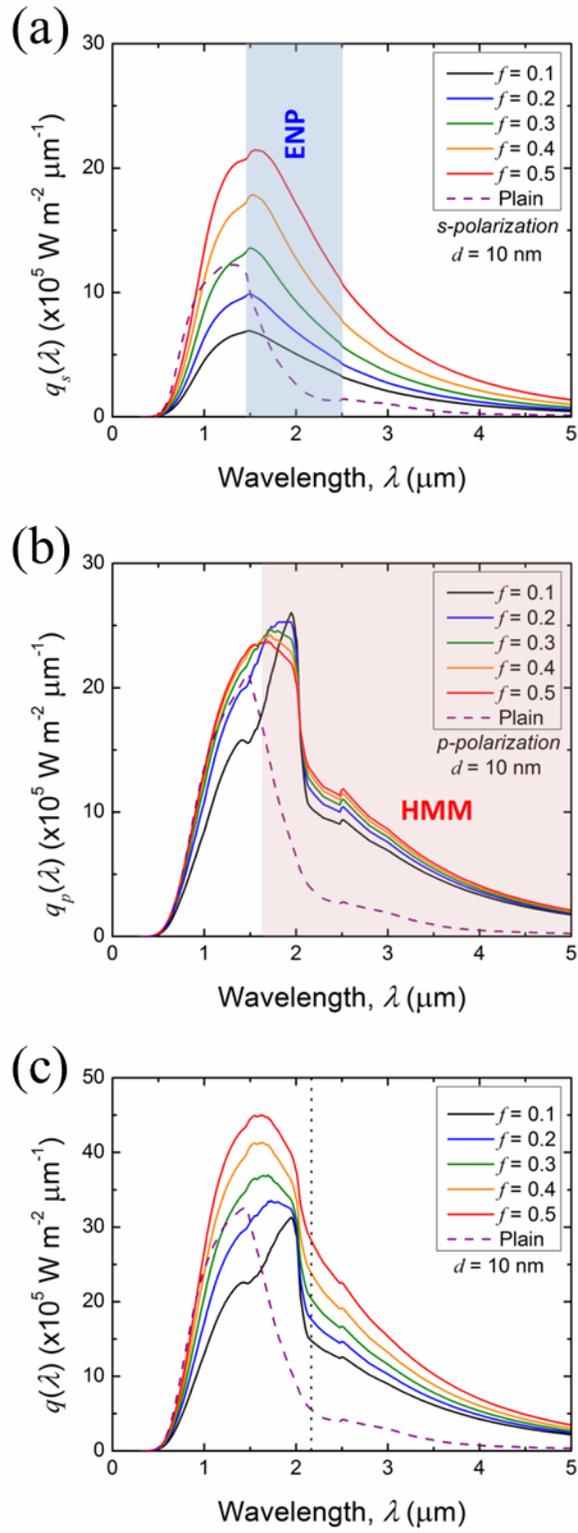

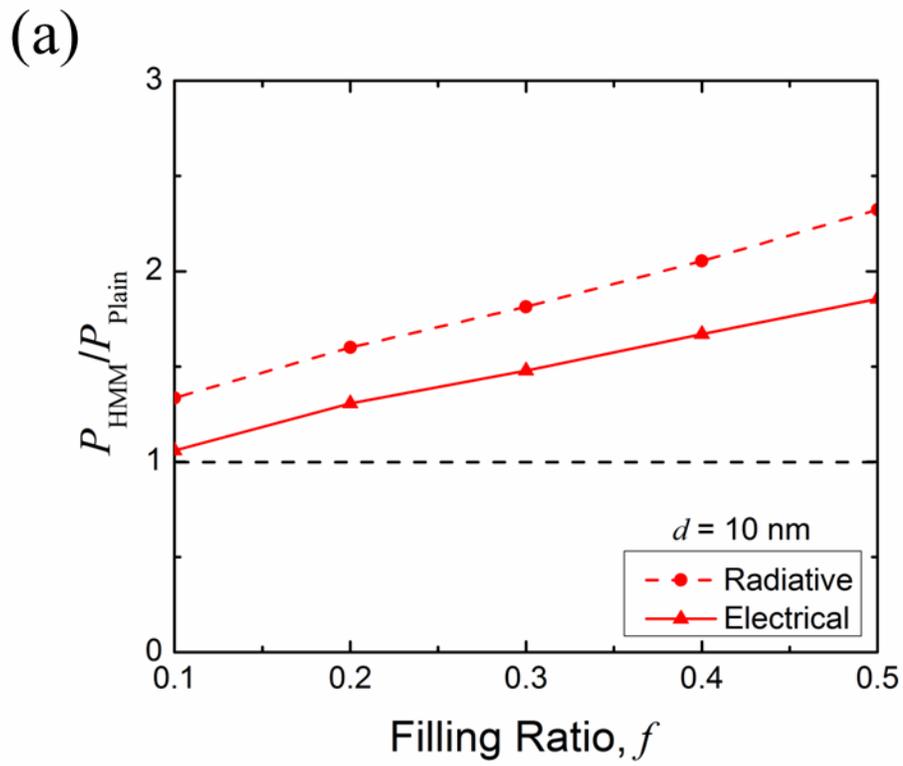

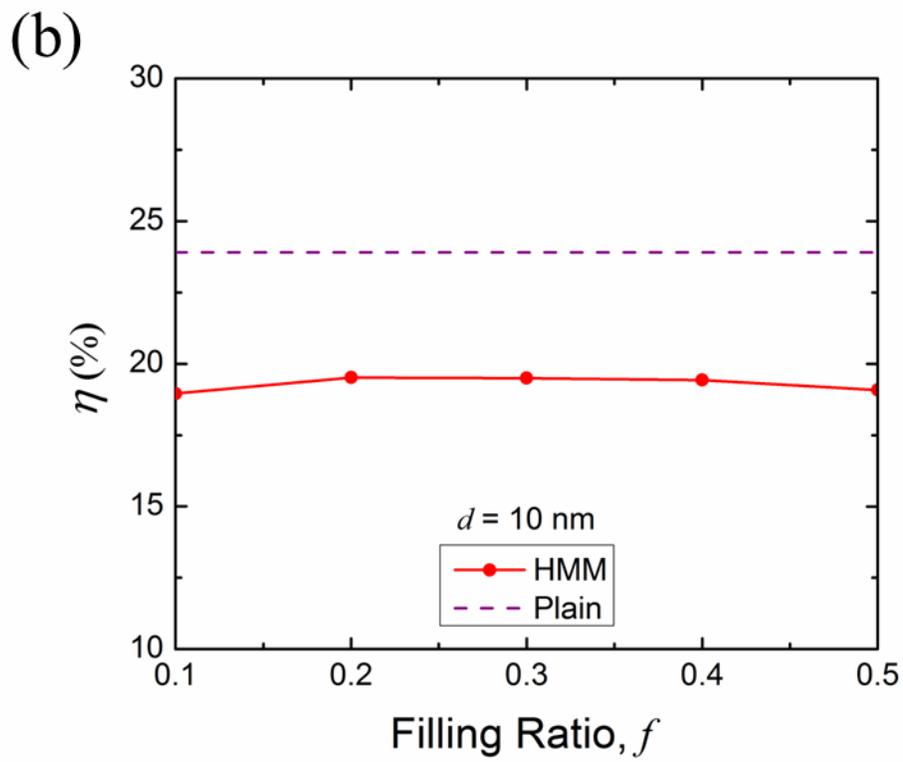

**Chang et al, Fig. 5**

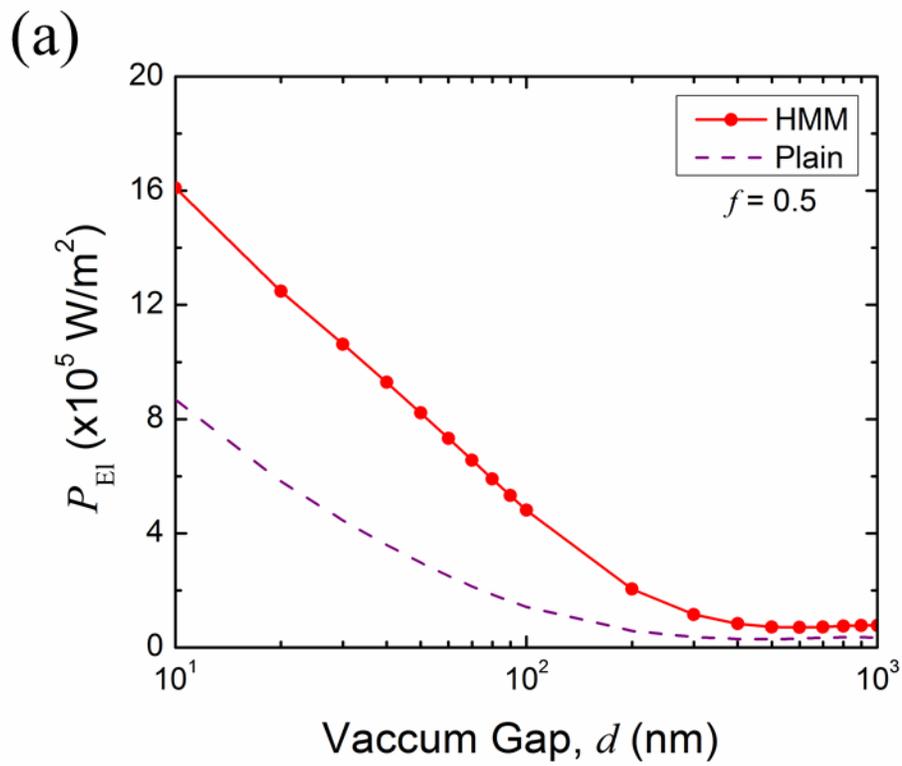

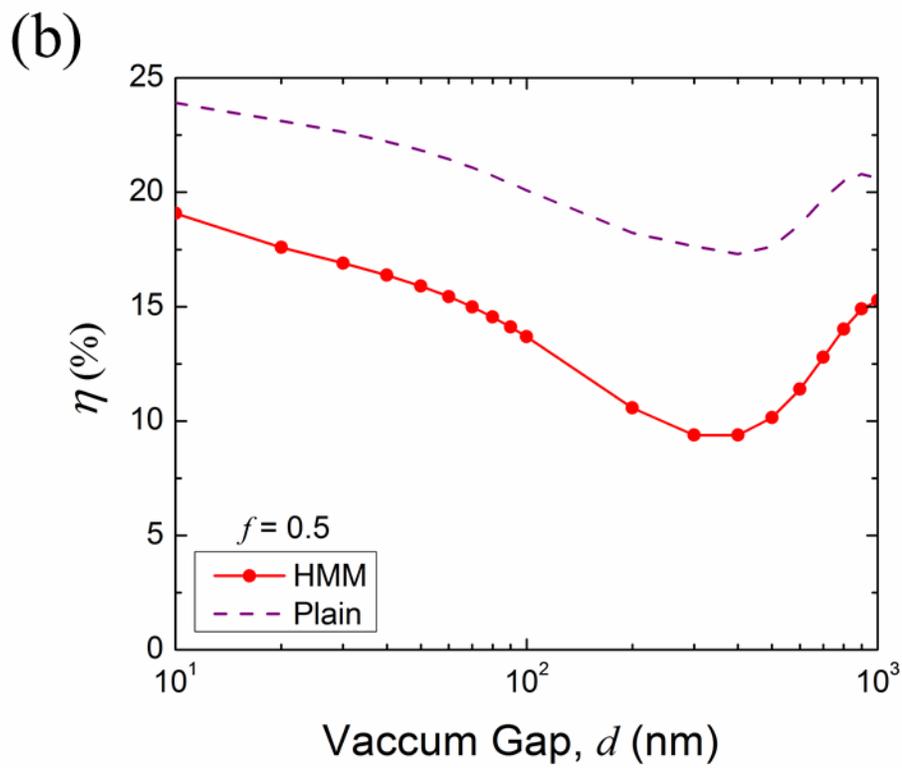

**Chang et al, Fig. 6**

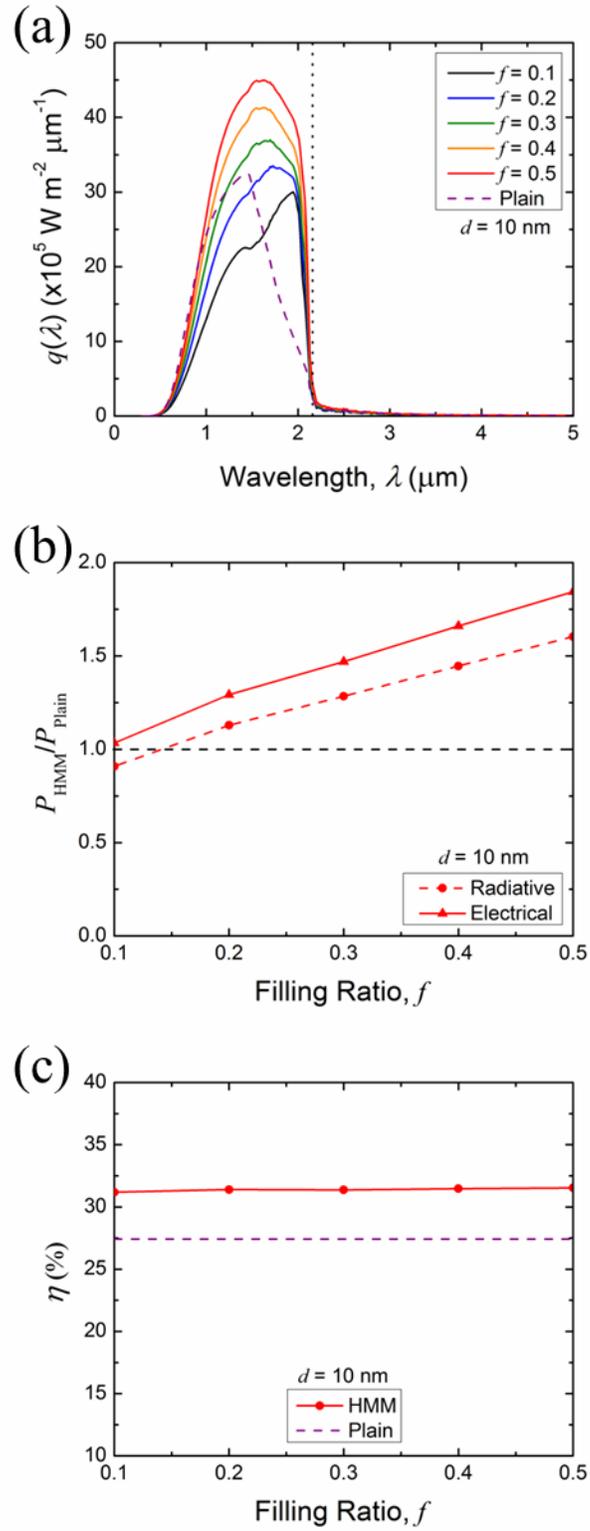

**Chang et al, Fig. 7**

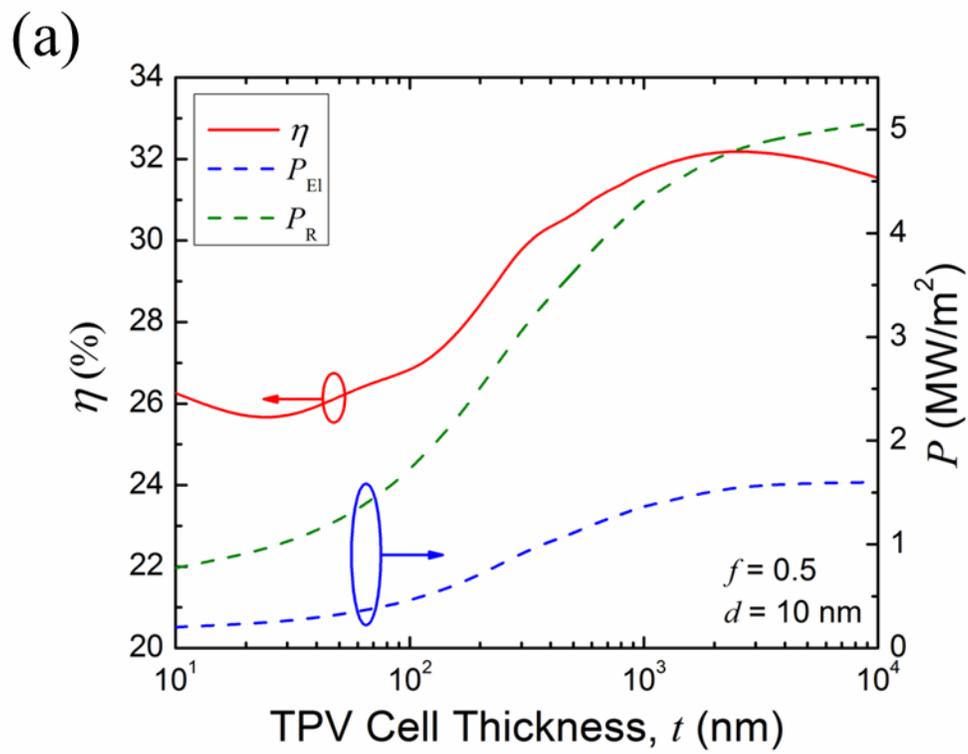

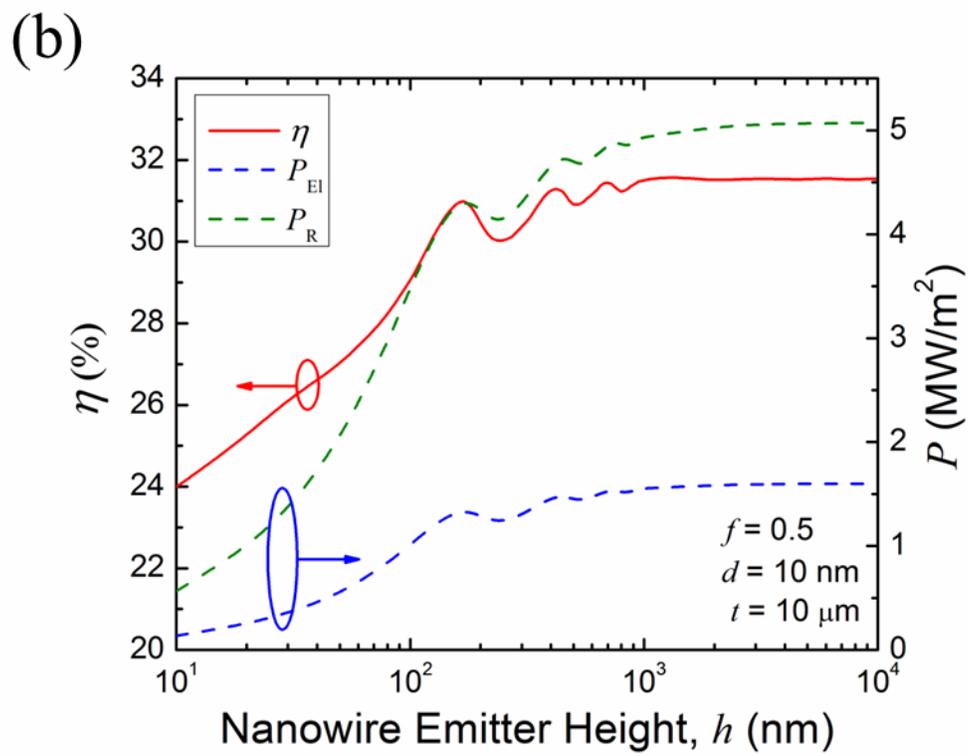

**Chang et al, Fig. 8**